\documentstyle[12pt]{article}
\begin{document}
\thispagestyle{empty}
\def\cqkern#1#2#3{\copy255 \kern-#1\wd255 \vrule height #2\ht255 depth
   #3\ht255 \kern#1\wd255}
\def\cqchoice#1#2#3#4{\mathchoice%
   {\setbox255\hbox{$\rm\displaystyle #1$}\cqkern{#2}{#3}{#4}}%
   {\setbox255\hbox{$\rm\textstyle #1$}\cqkern{#2}{#3}{#4}}%
   {\setbox255\hbox{$\rm\scriptstyle #1$}\cqkern{#2}{#3}{#4}}%
   {\setbox255\hbox{$\rm\scriptscriptstyle #1$}\cqkern{#2}{#3}{#4}}}
\def\CC{\mathord{\cqchoice{C}{0.65}{0.95}{-0.1}}}
\def\x{\stackrel{\otimes}{,}}
\def\y{\stackrel{\circ}{\scriptstyle\circ}}
\def\proof{\noindent Proof. \hfill \break}
\def\a{\begin{eqnarray}}
\def\b{\end{eqnarray}}
\def\o{\overline}
\def\p{{1\over{2\pi i}}}
\def\Q{{\scriptstyle Q}}
\def\P{{\scriptstyle P}}
\def\t{\textstyle}
\begin{flushright} ENSLAPP-L-422/93\\
 JINR E2-93-72
 \end{flushright}
\vskip0.5cm
\centerline{\LARGE N=2 Superconformal Affine Liouville Theory}
\vspace{1truecm}
\vskip0.5cm
\centerline{{\large E. Ivanov}}
\vskip.5cm
\centerline{Bogoliubov Theoretical Laboratory, JINR,}
\centerline{Dubna, Head Post Office, P.O. Box 79,}
\centerline{101000, Moscow, Russia}
\vskip1cm
\centerline{{\large F. Toppan}}
\vskip.5cm
\centerline{Laboratoire de Physique Th\'{e}orique ENSLAPP,}
\centerline{Ecole
Normale Sup\'erieure de Lyon,}
\centerline{46 All\'ee d'Italie, 69007 Lyon, France}
\vskip.5cm
\vskip1.5cm
\centerline{\bf Abstract}

\vskip.5cm
We present a new supersymmetric integrable
model: the $N=2$ superconformal affine Liouville theory.
It interpolates between the
$N=2$ super Liouville and $N=2$ super sine-Gordon theories and
possesses a Lax representation on the complex affine
Kac-Moody superalgebra ${\hat {sl(2| 2)^{(1)}}}$.
We show that the higher spin $W_{1+\infty}$-type symmetry algebra
of ordinary
conformal affine Liouville theory extends to a
$N=2\; W_{1/2 + \infty}$-type superalgebra.
\newpage\setcounter{page}1

\section{Introduction}

It is a well-established fact [1-4]
that an important class of
conformally
invariant integrable models can be produced starting from constrained
WZNW theories
based on affine Kac-Moody algebras (Conformally Affine Toda or CAT theories).
The standard massive affine Toda
theories can be
recovered
as a special limit of these models.

At the bosonic level any given simple Lie algebra admits
an affine extension and the associate CAT model can be defined, with a
Lax pair given on this affine extension.

In the supersymmetric case this is not always possible: to obtain a
manifestly
supersymmetric integrable model one needs to start from superalgebras
admitting a set of purely fermionic simple roots, the reason for that being
the
fact that the Lax pair operators should be fermionic objects in this case.
So, to learn
which superalgebras admit an $N=1$ superextended integrable CAT model it is
sufficient to look at the classification of superalgebras and
their root systems
as is given in \cite{FSS}.  For instance, the $N=1$ superconformal
affine Liouville (super CAL) theory \cite{TZ} which generalizes the
massive super sinh-Gordon model \cite{OBL} is associated with the twisted
affine superalgebra ${\hat { {osp(2|2)}^{(2)}}}$ having two fermionic simple
roots.

As for the $N=2$ case, the demand that a superalgebra admits a set of purely
fermionic simple roots is necessary but not sufficient; a detailed discussion
of
which superalgebras can give rise to integrable models possessing a
second supersymmetry has been given in [7-9]. The simplest
$N=2$ superconformal integrable model is the $N=2$ super Liouville theory
\cite{IK}
associated with the superalgebra $sl(2|1)$ and the simplest $N=2$ massive
integrable model is the $N=2$ super sine-Gordon theory associated
with the loop superalgebra $sl(2|2)^{(1)}$ \cite{{EH2},{IKKU}}.

In this letter we present an integrable $N=2$ superconformal affine Liouville
model which reduces to
the $N=2$ super sine-Gordon and the $N=2$ super Liouville models in
two special limits.  We will work in a manifestly
supersymmetric $N=2$ superfield approach. At first we will introduce our
model in the lagrangian
formulation and then we will discuss it in a Lax pair context \cite{LS}.
We will also
construct supercurrents which generate an infinite-dimensional
higher spin symmetry superalgebra generalizing the $W_{1+\infty}$
type symmetry algebra of the ordinary CAT models \cite{AFGZ2}.

\section {The action and equations of motion}

We will work in a manifestly supersymmetric formalism.

Let $x^{\pm\pm}$ denote the
bosonic $2D$ light-cone coordinates,
then the $N=2\;2D$ superspace is parametrized by $x^{\pm\pm}$ and the complex
fermionic coordinates
$\theta^\pm$ (and their conjugate ${\overline{\theta}^\pm}$). The $N=1$
superspace is recovered by letting ${{\theta^\pm}}={\overline{\theta}^\pm}$.
The $N=2$ spinor derivatives $D_{\pm},{\overline D}_{\pm}$ are
defined as:
\a
D_{\pm} &=&\textstyle{\partial\over{\partial\theta^{\pm}}}-
i{\overline \theta}^{\pm}\partial_{\pm\pm}\nonumber\\
{\overline D}_{\pm} &=&-{\t{\partial\over\partial{\overline\theta}^{\pm}}}+
i{ \theta}^{\pm}\partial_{\pm\pm} \;.
\b
The only non-vanishing bracket between them is given by
\a
\{D_\pm, {\overline D}_\pm \} &=& 2i \partial_{\pm\pm} \;.\nonumber
\b
In particular we have
\a
{D_\pm}^2 = {{\overline D}_{\pm}}^2 =0&& \;.\nonumber
\b
An $N=2$ chiral superfield $\Psi$, the simplest matter $N=2$ supermultiplet,
can be defined by the constraint
\a
{{D}_\pm}\Psi = 0, &&\nonumber
\b
while its conjugate satisfies
\a
{\overline D}_\pm \Psi^{\dagger} =0 \nonumber
\b
and so it is an anti-chiral $N=2$ superfield.

The bosonic conformal affine Liouville theory is formulated in terms of
three real bosonic fields \cite{BB}. A natural way to promote it to a
$N=2$ supersymmetric theory is to put these fields in proper minimal
$N=2$ supermultiplets. So we will define the $N=2$ superconformal affine
Liouville theory via three $N=2$ chiral superfields
$\Phi,\;\Lambda,\;\Sigma$.
By analogy with the $N=0$ and $N=1$ \cite{{BB},{TZ}} cases we choose the
action ${\cal S}$ to be
\a
{\cal S}&=& {\textstyle {1\over 4}} \int d^2 x d^2{\theta_+} d^2{\theta_-}
\{ \Phi
{\Phi^\dagger}+\Sigma{\Lambda^\dagger}+\Lambda{\Sigma^\dagger}+\nonumber\\
&&
\alpha e^{\Phi}\theta_+\theta_-+\beta e^{\Lambda-\Phi}\theta_+\theta_-
+\alpha^{\dagger} e^{\Phi^\dagger}{\overline\theta}_-{\overline\theta_+}+
\beta^{\dagger} e^{\Lambda^{\dagger}
-\Phi^{\dagger}}{\overline\theta}_-{\overline\theta}_+
\}\;.
\label{action1}
\b
Without loss of generality, the constants $\alpha ,\;\beta $ can be chosen
equal
unity, $\alpha ,\;\beta =1$ (by means of proper constant shifts of the
superfields $\Phi ,\;\Lambda$ and their conjugates).

{}From the above action the following superfield equations of motion can be
derived:
\a
{\overline D}_+{\overline D}_- \Phi &=& -e^{\Phi^{\dagger}} +
e^{\Lambda^{\dagger} -\Phi^{\dagger}} \nonumber\\
{\overline D}_+{\overline D}_- {\Sigma } &=& - e^{\Lambda^{\dagger} -
\Phi^{\dagger}}\nonumber\\
{\overline D}_+
{\overline D}_- \Lambda &=& 0
\label{eqmo}
\b
(and their conjugates). The $N=2$ super sine-Gordon theory
\cite{{EH2},{IKKU}} is
recovered as the special
solution
$\Lambda = \Lambda^\dagger = 0$. On the other hand, redefining $\Lambda$ as
$\Lambda \rightarrow a\Lambda$ and letting $a \rightarrow \infty $, one
recovers
the $N=2$ super Liouville theory \cite{IK}. In the bosonic limit the
theory constructed
becomes a conformal affine generalization of both the
sine- and sinh-Gordon theories, as it is evident from the equations of
motion
written in terms of the component fields (see below).

Let us introduce the component fields as follows:
\a
\varphi &\equiv& \Phi| \nonumber\\
\psi_+ &\equiv& {\o D}_+\Phi| \nonumber\\
F &\equiv& {\o D}_+{\o D_-}\Phi| \nonumber\\
\lambda &\equiv& \Lambda| \nonumber\\
\mu_+ &\equiv& {\o D}_+\Lambda| \nonumber\\
L &\equiv& {\o D}_+{\o D_-}\Lambda| \nonumber\\
\sigma &\equiv& \Sigma| \nonumber\\
\rho_+ &\equiv& {\o D}_+\Sigma| \nonumber\\
S &\equiv& {\o D_+}{\o D_-}\Sigma|\;.
\b
All these fields are complex, $\varphi ,\;\lambda ,\; \sigma$ being bosonic,
$\psi_+ ,\;\mu_+ , \;\rho_+ $ fermionic and
$F,\; L,\; S $ auxiliary.

After eliminating the auxiliary fields by their equations of motion we are
led
to the following expression for the action $\cal S$ in terms of
component fields:
\a
{\cal S}&=& \int d^2x\{\partial_{++}\varphi\partial_{--}\varphi^\dagger +
\partial_{++}
\sigma\partial_{--}\lambda^\dagger +
\partial_{++}\lambda\partial_{--}\sigma^\dagger \nonumber\\
&&
-{\textstyle{i\over 2}} \partial_{--}\psi_+{\o\psi}_+ -
{\textstyle{i\over 2}} \partial_{++}\psi_-{\o\psi}_-
-{\textstyle{i\over 2}} \partial_{--}\rho_+{\o\mu}_+
-{\textstyle{i\over 2}}
\partial_{++}\rho_-{\o \mu}_-
\nonumber\\
&&
 -{\textstyle{i\over 2}} \partial_{--}\mu_+{\o\rho}_+ -
{\textstyle{i\over 2}} \partial_{++}\rho_-{\o\mu}_-
-{\textstyle{1\over 4}} \psi_-\psi_+e^{\varphi}+
{\textstyle{1\over 4}} {{\o\psi}_-}
{\o\psi}_+e^{\varphi^\dagger}
\nonumber\\
&&
-{\textstyle{1\over 4}} (\mu_--\psi_-)(\mu_+-\psi_+)e^{\lambda-\varphi}+
{\textstyle{1\over 4}} ({\o\mu}_--{\o\psi}_-)({\o \mu}_+-{\o \psi}_+)
e^{\lambda^\dagger-\varphi^\dagger}\nonumber\\
&&-{\textstyle{1\over 4}} e^{\varphi+ \varphi^{\dagger}} -
{\textstyle {1\over 4}}
e^{\lambda - \varphi} e^{\lambda^{\dagger} - \varphi^{\dagger}}
+ {\textstyle{1\over 4}} e^{\lambda -\varphi} e^{\varphi^{\dagger}}  +
{\textstyle{1\over 4}} e^{\lambda^{\dagger} -
\varphi^{\dagger}} e^{\varphi} \}\;.
\b
The equations of motion for the bosonic fields are:
\a
& \partial_{++}\partial_{--}\varphi -
{1\over 4}{\o\psi}_-{\o\psi}_+e^{\varphi^\dagger} +{1\over 4}
({\o\mu}_--{\o\psi}_-)({\o\mu}_+-{\o\psi}_+)
e^{\lambda^\dagger -\varphi^\dagger} &\nonumber\\
&+{1\over 4}
e^{\varphi^\dagger +\varphi } - {1\over 4}e^{\lambda^\dagger +\lambda }
e^{-(\varphi^\dagger +\varphi) }
- {1\over 4} e^{\lambda +\varphi^\dagger-\varphi} +
{1\over 4}e^{\lambda^\dagger-\varphi^\dagger +\varphi} =0&\nonumber\\
&\partial_{--}\partial_{++}\sigma -{1\over 4}({\o\mu}_- -{\o\psi}_- )
({\o\mu}_+
- {\o\psi}_+ )
e^{\lambda^\dagger -\varphi^\dagger} + {1\over 4}
e^{\lambda^\dagger +\lambda +\varphi-\varphi^\dagger} -
{1\over 4} e^{\lambda^\dagger +\varphi-\varphi^\dagger}=0&\nonumber\\
&\partial_{--}\partial_{++}\lambda =0\;.&
\b
In the bosonic limit, with all fermions discarded, they are reduced to the
system
\a
\partial_{--}\partial_{++} (Re\varphi ) +{\textstyle {1\over 4}}e^{2Re\varphi}
-{\textstyle {1\over 4}}e^{2Re\lambda - 2 Re\varphi} &=&0\nonumber\\
\partial_-\partial_+ (Im\varphi ) -{\textstyle {1\over 2}}e^{Re\lambda}
sin(Im\lambda -2Im\varphi ) &=& 0
\nonumber\\
\partial_{--}\partial_{++} (Re\sigma ) +{\textstyle {1\over 4}}e^{2Re\lambda}
cos( 2Im\varphi )
-{\textstyle {1\over 4}}e^{Re\lambda}
cos(Im\lambda - 2 Im \varphi )&=&0\nonumber\\
\partial_{--}\partial_{++} (Im\sigma ) +{\textstyle {1\over 4}}e^{2Re\lambda}
sin(2Im\varphi )
+{\textstyle {1\over 4}}e^{Re\lambda }sin( Im\lambda-2Im\varphi)
&=&0\nonumber\\
\partial_{--}\partial_{++} (Re\lambda) =
\partial_{--}\partial_{++} (Im\lambda )  &=&0\;.
\label{action2}
\b
As was mentioned above, this set is a conformally invariant extension of
both the sinh- and sine-Gordon equations (for the fields
$Re \phi $ and $Im \phi $, respectively) which are restored in the limit
$\lambda = 0$.

We end this section by giving the equations of motion for
the fermionic fields
\a
\partial_{--} \psi_+ +
{\textstyle{i\over 2}}{\o\psi}_- e^{\varphi^\dagger} - {\textstyle{i\over 2}}
({\o\mu}_--{\o\psi}_- ) e^{\lambda^\dagger -\varphi^\dagger} &=&0
\nonumber \\
\partial_{++}\psi_- -{\textstyle{i\over 2}}{\o\psi}_+ e^{\varphi^\dagger} +
{\textstyle{i\over 2}}({\o\mu}_+-{\o\psi}_+ )
e^{\lambda^\dagger -\varphi^\dagger} &=&0
\nonumber \\
\partial_{--} \rho_+ +
{\textstyle{i\over 2}}({\o\mu}_--{\o\psi}_- )
e^{\lambda^\dagger -\varphi^\dagger} &=& 0
\nonumber \\
\partial_{++} \rho_- + {\textstyle{i\over 2}} ({\o\mu}_+-{\o\psi}_+ )
e^{\lambda^\dagger -\varphi^\dagger} &=& 0
\nonumber \\
\partial_{--} \mu_+ = \partial_{++}\mu_-
 &=& 0\;.
\b
\section{Lax pair formulation}

The model defined in the previous section
is the conformal affine extension of the
$N=2$ super sine-Gordon theory considered in ref. \cite{{EH2},{IKKU}}.
The latter
theory
is integrable: its
equations can be cast in a Lax form with the Lax
connections
valued in the loop superalgebra $sl(2|2)^{(1)}$.
In this section we will make explicit
the integrability properties
of our model by writing down its Lax pair formulation, which turns out
to be based on an affine extension of $sl(2|2)^{(1)}$.

Before going on, let us first recall some basic facts about $sl(2|2)^{(1)}$
\cite{{FSS},{EH2}}.
This superalgebra contains four bosonic generators $h_i$ in the Cartan
subalgebra and a set
of four simple roots generators
${e^{\pm}}_i$
(these all are fermionic). The
(anti)commutation relations
between the Cartan and simple roots generators are given by
\a
[ h_i, {e^{\pm}}_j ]&=& \pm a_{ij} {e^{\pm}}_j\nonumber \\
\{ {e^+}_i,{e^-}_j\}&=& \delta_{ij}h_j\;,
\b
with $ i,j =1,...,4$ and $ a_{ij}$ being the Cartan matrix
\a
a_{ij} &=& \left( \begin{array}{cccc}
0 & -1 & 0 & 1\\
-1 & 0 & 1 & 0\\
0&1&0&-1\\
1&0&-1&0
\end{array}\right)\;.
\b
The above commutation relations are the only ones which are actually
needed to completely define the massive integrable theory corresponding to
$sl(2|2)^{(1)}$.

The key property of the above Cartan matrix is that it is degenerate, having
rank $r=2$: the Cartan elements $h_1+h_3$ and $h_2+h_4$ are
$c$-numbers which commute with all other  generators.

This fact is an indication that the
integrable theory associated with this superalgebra, i.e. the
$N=2$ super sine-Gordon
theory, is not conformally invariant:
it is already known from the bosonic \cite{{BB},{BMZ}} and $N=1$
supersymmetric \cite{TZ} cases
that conformally invariant integrable models can be produced
only starting
from (super)algebras which have
enough Cartan generators to remove the degeneracy
among
different roots.
The best example is supplied by the
sinh-Gordon theory related to the
$sl(2)$-loop algebra. A conformally invariant generalization of such a model
can be constructed
by enlarging the $sl(2)$-loop algebra with additional new generators in
its Cartan sector (a central extension and a
derivative operator counting the powers of the loop spectral parameter):
as a result one goes from the $sl(2)$-loop algebra to its affine extension.

A naive transfer of this procedure to the case of $N=2$ super
CAL theory does not work basically
for two reasons: first, because the introduction
of the derivative operator cannot completely remove the degeneracy
among roots (an additional operator is needed for
this purpose) and, second, because such a derivative is a real
operator while we are dealing with complex superfields in the $N=2$ case.

An essentially new feature of the superalgebra $sl(2|2)^{(1)}$ compared to
the (super)algebras relevant to the previously known
conformal affine theories consists in the existence of an
involution between its generators. Just this feature allows for the
second supersymmetry. One should therefore expect that the extension of
$sl(2|2)^{(1)}$ appropriate to the present case is
realized via the introduction of a pair of operators in involution.
Indeed, our model involves chiral superfields
and their conjugates.

Let us discuss in detail how such an extension can be constructed.

At first we notice that the involution just mentioned is
realized, in the particular
case of Cartan and simple roots generators, as
\a
&&
\begin{array}{cc}
h_1 \leftrightarrow h_2 & h_3\leftrightarrow h_4 \\
{e^+}_1\leftrightarrow {e^+}_2 & {e^+}_3 \leftrightarrow {e^+}_4\;.
\end{array}
\b
Since no confusion will arise, let us call for simplicity from now on
the above involution as ``conjugation" operation.

To enlarge the superalgebra we introduce a couple
of conjugate bosonic operators $d,{\o d}$. These are defined as follows:
the commutator between them and a generator $e_\alpha$ corresponding to
the root $\alpha$,
with the decomposition $\alpha =\sum_in_i\alpha_i$ for $i=1,...,4$,
$n_i$ integer and $\alpha_i$ denoting the simple roots,
is given by:
\a
\relax [d, e_\alpha ] &=& -n_4e_\alpha \nonumber\\
\relax [{\o d}, e_\alpha ] &=& n_3e_\alpha \;,
\label{rootop}
\b
while their commutators with themselves
and with the Cartan generators $h_i$ are assumed to vanish.

Due to the properties of the root decomposition the above relations
are consistent with the Jacobi identities.

The operators ${\o d},\; { d}$ are grading operators with respect to the
roots ${e^+}_3$,
${e^+}_4$, respectively. The choice of grading with respect to
just these roots is to some extent arbitrary, the important point
is that the root operators should be conjugate. For standard integrable
theories
associated with non-singular Cartan matrices such grading operators
are not too useful notion since they can be re-expressed as combinations of
the original Cartan generators. The importance of $d,{\o d}$ in the case
at hand is that they allow to remove the degeneracy of the Cartan subalgebra.

We will denote the algebra enlarged by $d, {\o d}$ as
${\hat {sl(2|2)^{(1)}}}$:
$$
{\hat{sl(2|2)^{(1)}}}= sl(2|2)^{(1)}\oplus
{\bf C} d\oplus {\bf C} {\o d}
$$
Clearly, $d,\; {\o d}$ belong to the Cartan sector of
${\hat{sl(2|2)^{(1)}}}$.

It should be mentioned that, in the bosonic and $N=1$ supersymmetric cases
[1,4], in order to obtain a conformally invariant theory one is forced to
introduce
not only the derivative operator, but also the central extension. In the
present case we do not need to further enlarge the algebra with central
extensions, because we have already at our disposal a couple of
(conjugate) $c$-numbers given by $h_1+h_3$, $h_2+h_4$ which play such a role.

To express our formulas in a concise notation, it is convenient to split
the Cartan subalgebra in the two conjugate parts
\a
{\cal H}&=& \{h_1, h_3, d\}\nonumber\\
{\o{\cal H}}&=& \{ h_2, h_4, {\o d} \}
\b
and to denote the sum over the sets of two conjugate simple roots by
\a
&& {\cal E}^{\pm} = {e^\pm}_1 +{e^\pm}_3\quad\quad
{\o{\cal E}}^{\pm} = {e^\pm}_2+{e^\pm}_4\;.
\b
Further, let us denote with $\Psi , \;{\o \Psi}$ two conjugate superfields
taking values in the Cartan subsectors ${\cal H},\; {\o {\cal H}}$,
respectively:
\a
\Psi &=&    A^\dagger h_1 + B^\dagger h_3 + C^\dagger d \nonumber\\
{\o\Psi} &=& A h_2 + Bh_4 + C {\o d} \;.
\b
We possess now all the necessary ingredients to express the integrability
properties
of our system by introducing two conjugate linear systems of the Lax pair
type:
\begin{equation}
(D_\pm+{\cal L}_\pm){\cal T}=0
\end{equation}
and
\begin{equation}
({\overline D}_\pm+{\overline{\cal L}}_\pm){\overline {\cal T}}=0\;,
\end{equation}
where ${\cal L}_\pm, {\o{\cal L}}\pm$ belong to the superalgebra
${\hat{sl(2|2)}^{(1)}}$ and ${\cal T}, {\o{\cal T}}$
to its associate affine KM supergroup.

The zero-curvature condition is the compatibility condition of the
above linear systems and it is provided by the relations
\a
D_+{\cal L}_-+D_-{\cal L}_++\{{\cal L}_+,{\cal L}_-\}&=&0\nonumber\\
{\overline D}_+{\overline {\cal L}}_- + {\overline D}_-{\overline {\cal L}}_++
\{{\overline {\cal L}_+},{\overline {\cal L}_-}\} &=&0\;.
\b
To ensure integrability, the equations of motion of the lagrangian
formulation (3)
should be recovered from the above zero-curvature relations.
This is indeed the case.

Let us define
\a
L_+ &=& D_+\Psi + e^{ad {\o \Psi} }{\cal E}^+
\nonumber\\
L_- &=&  -{\cal E}^-
\b
for one copy of the Lax pair, and
\a
{\o L}_+ &=& - {\o{\cal E}}^+\nonumber\\
{\o L}_- &=& {\o D}_- {\o\Psi} + e^{ad\Psi}{\o{\cal E}}^-
\b
for the conjugate copy.
Explicitly we have, in the former case,
\a
L_+ &=& (D_+{A}^\dagger) h_1 +(D_+{B}^\dagger) h_3 + (D_+ {C}^\dagger) d
+\nonumber\\
&&  e^{(B-A)}{e^+}_1 + e^{(C-(B-A))} {e^+}_3 \nonumber\\
L_- &=&  -({e^-}_1+{e^{-}}_3)
\b
(and similar expressions in the latter case).
It is a straightforward exercise to
check that the zero-curvature condition is satisfied, once
provided that the
superfields $A,B,C$ and their conjugates satisfy the following
relations
\a
D_+D_-A^{\dagger} &=& -e^{(B-A)}\nonumber\\
D_+D_- B^\dagger &=& -e^{(C-(B-A))}
\nonumber\\
D_+D_- C^\dagger &=& 0
\label{Laxmotion}
\b
together with their conjugate counterparts.

It is worth mentioning that the zero appearing
in the r.h.s. of the equation for the
superfield $C^\dagger$ is due to the fact that the $d$ generator
never appears in the commutators of
generators of $sl(2|2)^{1}$.
Moreover, in the r.h.s. of all equations the superfields $A,B$ come out only
in the combination
$B-A$ since $h_1+h_3$ is a $c$-number.

The Lax-pair version of the theory involves $3$ complex superfields
just like the lagrangian formulation of the previous section.
The zero-curvature
equations (\ref{Laxmotion})
reproduce just
the equations of motion of the lagrangian theory under the following
identification:
\a
\Phi\equiv B-A,\quad \Lambda \equiv C, \quad \Sigma \equiv -B
\b
(the resulting equations are conjugate of eqs. (3)).

\section{Superconformal properties and higher spin primary fields}

A characteristic feature of the model we have constructed is its invariance
under
$N=2$ superconformal transformations which are infinitesimally given by
\a
&&
\begin{array}{cc}
\delta \Phi = -D_+\delta\theta^+ & \delta\Lambda =-2D_+\delta\theta^+ \\
\delta\Sigma =
\alpha D_+\delta\theta^+ & \delta D_+ = -(D_+\delta\theta^+)D_+ \\
{\o D}_+\delta \theta^+ = 0
\end{array}
\label{transf}
\b
and by their conjugate counterparts.
Here $\alpha$ is an arbitrary real parameter.
It is straightforward to check that the action (\ref{action1})
and the equations
of motion (\ref{eqmo})
are invariant with respect to these transformations.
Similar relations hold for the superconformal transformations acting
in the opposite light-cone sector of $N=2$ superspace (on the
coordinates $x^{--}, \theta^-, {\o\theta}^-$).

The presence of extra superfields $\Lambda $, $ \Sigma $ allows to
reestablish the superconformal invariance which is spoiled
in the corresponding massive model ($N=2$ super sine-Gordon theory).
Due to these superfields the constructed model turns out to
have a richer algebraic
structure than both its special limits, the $N=2$ super Liouville and
super sine-Gordon theories. In particular, the action (\ref{action1})
is
invariant under the shifts
$$
\Sigma \rightarrow \Sigma + \delta \Sigma\;,\;\;\;
\Sigma^{\dagger} \rightarrow \Sigma^{\dagger} + \delta \Sigma^{\dagger}\;,
$$
with $\delta \Sigma$ ($\delta \Sigma^{\dagger}$) being a sum of two arbitrary
chiral (anti-chiral) superfunctions living in the left and right light-cone
sectors of $N=2$ superspace. By the way, using this freedom, one may always
redefine the superconformal transformation of $\Sigma$ so as to make it
homogeneous (this amounts to choosing $\alpha = 0$ in eqs. (24)).

Recall that the bosonic analog of our
model, the CAL theory (as well as its
Toda generalizations), exhibits a $W_{1+\infty}$ type symmetry
algebra generated by an infinite set of higher spin primary fields together
with the spin 2 conformal stress-tensor and a quasi-primary spin 1 field
\cite{AFGZ2}. We will repeat in our case the procedure developed
in \cite{AFGZ2} and will find an infinite set of supercurrents generating a
$N=2$ superextension of the algebra present in the CAL theory.
Like in most (super)conformally invariant theories, this superalgebra
splits into two commuting light-cone copies. So, without loosing generality,
we will limit our study
to one of them corresponding to the light-cone
direction specified by $x^{++} , \theta^+ ,{\o\theta}^+$.

Let us first recall the definition of primary $N=2$ superfields which are a
natural generalization of the spin $s$ primary fields used in the
construction of ref. \cite{AFGZ2}.
These are denoted by $T^{(s_1\;s_2)}$, with $s_1$, $s_2$ being arbitrary
integers, and transform under the $N=2$ superconformal group (\ref{transf})
according to the rule
\a
\delta T^{(s_1\;s_2)} & =& (s_1{\o D}_+\delta {\o\theta}^+ -
s_2D_+\delta\theta^+ )\cdot T^{(s_1\;s_2)}\;.
\label{primary}
\b
The external conformal spin of $T$ is defined as $s \equiv {1\over 2}
(s_1 + s_2)$ while $h \equiv s_1 - s_2$ has the meaning of the external
$U(1)$ charge of $T$ (under the convention that the $U(1)$ charge of
$\theta^{\pm}$ equals 1).

The basic ingredient of the construction of ref. \cite{AFGZ2} is the anomaly
free spin 2 conserved
current $T_{++++}$. Its analog in the $N=2$ case is the anomaly
free conserved\footnote{A
supercurrent $J$
is called conserved if, by using the equations of motion for
the involved superfields, it
satisfies the relations
$D_-J = {\o D}_-J =0$ which imply in particular the condition of
light-cone chirality with respect to the ordinary space-time variables,
$\partial_{--} J=0$.}
spin 1 supercurrent $\tilde{J}_{++}^{(1\;1)}$. In order to find it we
proceed as follows.

As a first step we define the most general spin 1 real conserved
supercurrent which turns out to be
\a
J_{++} &=& {\o D}_+ \Phi D_+\Phi^\dagger + 2i\partial_{++}
(\Phi -\Phi^\dagger ) +\nonumber\\
&&
{\o D}_+ \Lambda D_+\Sigma^\dagger +{\o D}_
+\Sigma D_+\Lambda^\dagger +
4i \partial_{++} ( \Sigma - \Sigma^\dagger )+\nonumber\\
&& a
{\o D}_+ \Lambda D_+\Lambda^\dagger +
i b \partial_{++} (\Lambda-\Lambda^\dagger )\;,
\label{current}
\b
with $a,\; b$ being arbitrary real constants.

The parameters $a,\; b$ in (\ref{current}) are uniquely fixed
via the parameter $\alpha$ defined in (\ref{transf}) if we further
demand $J_{++}$ to
transform as a primary $(1,1)$ $N=2$
superfield, i.e. require the absence of an inhomogeneous piece in
its superconformal transformation law.
In this way we recover the expression for the anomaly-free supercurrent
${\tilde J}_{++}^{(1\;1)}$ which is given by eq. (\ref{current}) with the
following
fixed values of the involved parameters
\a
&& a= \alpha-{1\over 4} \quad\quad b = 2\alpha-1\;.
\b
Note that the canonical $N=2$ conformal supercurrent generating, via
appropriate super Poisson brackets,
superconformal transformations of the superfields $\Phi,\;\Lambda$
and $\Sigma$ in eqs. (26) corresponds to a different choice of
the parameters, namely,
\a
&& a=0 \quad \quad b = -2 \alpha\;.
\b
It is worth mentioning that it is anomaly-free at $\alpha = {1\over 4}$.

One more entity used in \cite{AFGZ2} to construct an infinite sequence
of the primary higher spin currents generating a $W$ type symmetry algebra
is a conserved spin 1 current. It is quasi-primary, i.e. inhomogeneously
transforms under the conformal group. $N=2$ analogs of this current
are two conjugate spin ${\t {1\over 2}}$ conserved supercurrents
$$
J_+ = D_+\Lambda^\dagger \;,\;\;\;\;{\o J}_+ ={\o D}_+\Lambda.
$$
These are quasi-primary with respect to superconformal transformations,
e.g.,
$$
\delta J_+ = - D_+\delta\theta^+ J_+ + 4i \partial_{++}\delta\theta^+\;.
$$

Now we are ready to generalize to our case the procedure
employed in \cite{AFGZ2}. We introduce the operators $I_1$, $I_2$
which count, respectively, the weights $s_1$ and $s_2$ of the primary
superfields
$$
I_{1,2} T^{(s_1\;s_2)} = s_{1,2} T^{(s_1\;s_2)}
$$
and define two supercovariant derivatives
\a
{\cal D}_+ &=& D_+ - {\textstyle{1\over 2}} D_+\Lambda^\dagger\; I_1\nonumber\\
{\o{\cal D}}_+ &=& {\o D}_+ - {\textstyle{1\over 2}}{\o D}_+ \Lambda\; I_2\;.
\b
They have the following properties: applied to a superfield
of weights $(s_1, s_2)$,
$T^{(s_1\;s_2)}$,
they send it, respectively,
into superfields of weights $(s_1, s_2+1)$ and
$(s_1+1,s_2)$. Indeed, it is a simple exercise to check that
the superfields
$$
{\cal D}_+ T^{(s_1\;s_2)}\;,\;\;\; {\o{\cal D}}_+ T^{(s_1\;s_2)}
$$
transform according to the generic transformation law (\ref{primary})
with the aforementioned weights.
Note that the supercovariant derivatives
${\cal D}_+$, ${\o{\cal D} }_+$
satisfy the following anticommutation relation:
\a
\{ {\cal D}_+, {\o{\cal D}}_+ \} &=& 2i\partial_{++}
+({\textstyle{1\over 4}} {\o D}_+ \Lambda D_+\Lambda^{\dagger}
-i\partial_{++}
\Lambda^{\dagger})I_1  +
({\textstyle{1\over 4}} D_+\Lambda^{\dagger} {\o D}_+\Lambda
-i\partial_{++}\Lambda^{\dagger})I_2  \nonumber \\
&& -{\textstyle{1\over 2}}D_+ \Lambda^{\dagger}
{\o D}_+ (I_1 +1) -{\textstyle{1\over 2}}{\o D}_+ \Lambda D_+(I_2 +1)
\nonumber\\
{\cal D}_+ {\cal D}_+ &=& {\o {\cal D}}_+{\o {\cal D}}_+ = 0\;.
\b

Now, starting with the anomaly-free spin 1 supercurrent
$\tilde{J}_{++}^{(1\;1)}$ and
acting on it successively by ${\cal D}$ and ${\o {\cal D}}$, we may
construct an infinite tower of conserved supercurrents with higher
$s_1$ and $s_2$ which is a genuine $N=2$ generalization of the
set of currents of the bosonic CAL model. All these are primary with
respect to the canonical $N=2$ conformal supercurrent,
i.e. have
no anomalous terms in their superconformal transformations. The generic form
of the basic supercurrents is as follows
(taking account of the relations (30))
\a
J_{+2n}^{(n\;n)} &=& {\o{\cal D}}_+ {\cal D}_+\; ....\;{\o{\cal D}}_+ {\cal
D}_+
\tilde{J}_{++}^{(1\;1)} \nonumber \\
J_{+(2n+1)}^{(n\;n+1)} &=& {\cal D}_+ J_{+2n}^{(n\;n)}\;.
\b
The remaining primary supercurrents are obtained from these two basic
sequences
via complex conjugation and permutation $s_1 \leftrightarrow s_2$.

Together with the spin ${\textstyle{ 1\over 2}}$ conserved supercurrents and
the canonical conformal supercurrent,
the primary supercurrents defined above form a set which is nonlinearly
closed under the super Poisson brackets and so is recognized as a kind
of infinite-dimensional nonlinear $N=2$ $W_{1/2 + \infty}$ superalgebra.
Based on analogy with the bosonic case \cite{AFGZ2},
in the limit of ``large'' $\Lambda$ these supercurrents are expected
to close on a $N=2$ supersymmetric
extension of the linear area-preserving $w_{1+\infty}$ algebra.
The detailed discussion will be reported elsewhere.

\section{Concluding remarks}

To summarize, in this paper we have constructed the $N=2$ conformal affine
super Liouville theory, shown its integrability by defining the appropriate
superfield Lax pair (zero curvature) representation and found an infinite
set of primary $N=2$ supercurrents which are counterparts of the analogous
currents of the bosonic CAT theories and form a nonlinear
$N=2\; W_{1/2 + \infty}$ type symmetry algebra of the model. It is
straightforward to extend this construction to arbitrary $N=2$ super Toda
theory \cite{EH} like this has been done for the bosonic case in ref.
\cite{AFGZ2}. An interesting question is how to reproduce $N=2$ CAT theories
from the appropriate supergroup $N=2$ supersymmetric WZNW sigma models via
hamiltonian reduction (for the bosonic and $N=1$ supersymmetric CAT theories
this has been done in \cite{AFGZ2} and \cite{TZ}). A trouble here is that
up to now a manifestly $N=2$ supersymmetric superfield formulation of
such sigma models for general target (super)groups is lacking. One more
interesting problem we are planning to address in the nearest future is the
construction of $N=4$ supersymmetric conformal affine Liouville theory
which should be an extension of $N=4$ super Liouville theory \cite{IK2}.

\vskip1cm
{\Large {\bf Acknowledgements}}

The authors are grateful to F. Delduc, L.A. Ferreira,
J.F. Gomes, S. Krivonos
for useful discussions. A special thank to P. Sorba
who explained us ref. \cite{FSS} and who carefully read the manuscript.
F.T. is grateful to Director of Bogoliubov Theoretical Laboratory of
JINR, Prof. D.V. Shirkov, for his kind
hospitality in Dubna, where part of this work has been
accomplished.

\end{document}